\documentclass[journal,twocolumn]{IEEEtran}

\usepackage{amsmath,amssymb,amsfonts}
\usepackage{graphicx}
\usepackage{algorithmic}
\usepackage{algorithm}
\usepackage{booktabs}
\usepackage{cite}
\usepackage{hyperref}
\usepackage{multirow}
\usepackage{xcolor}
\usepackage{amsthm}
\newtheorem{definition}{Definition}
\graphicspath{{./}}

\DeclareFontFamily{U}{stix2bb}{}
\DeclareFontShape{U}{stix2bb}{m}{n} {<-> stix2-mathbb}{}
\NewDocumentCommand{\indicator}{}{\text{\usefont{U}{stix2bb}{m}{n}1}}

\title{A Game-Theoretic Decision Framework for Optimal Selection of Coordination Detection Methods in Multi-UAV Fleet Operations}

\author{%
\IEEEauthorblockN{Christian G. Manasseh}
\IEEEauthorblockA{Mobius Logic Inc., Tysons, VA, USA}
\and
\IEEEauthorblockN{Savana J. Ammons}
\IEEEauthorblockA{Mobius Logic Inc., Tysons, VA, USA}
}

\begin{document}

\maketitle

\begin{abstract}
Detecting coordination among unmanned aerial vehicle (UAV) fleets operating in shared airspace and identifying the route-lead aircraft whose navigation decisions govern fleet behavior presents a fundamental speed--accuracy trade-off: fast methods enable real-time traffic management but sacrifice detection fidelity, while accurate methods may exceed the time budget for actionable airspace deconfliction.
This paper presents a \emph{game-theoretic decision framework} that resolves this trade-off by formulating method selection as a two-player zero-sum game between a Monitor (selecting computational methods and parameters) and Nature (selecting the unknown traffic scenario).
We construct an end-to-end pipeline from trajectory surveillance data through eight candidate detection algorithms, a Monte Carlo sensitivity analysis characterizing their stochastic performance, and finally a multi-objective optimization layer that identifies Pareto-optimal method portfolios.
The minimax solution provides a \emph{robust mixed strategy} with a probability distribution over methods that guarantees worst-case performance regardless of scenario uncertainty.
Experimental evaluation across 400 randomized configurations spanning 5--50 aircraft demonstrates that the framework recommends distinct method portfolios depending on operational priority. Graph Signal Processing dominates all operational priorities, but by different amounts: 59.9\% for the balanced profile, 58.5\% for the accuracy-priority, 65.4\% for the route-lead-priority, and 100\% for the speed-priority. 
The framework achieves a guaranteed game value of 0.20–0.45 (normalized utility) across all tested preference profiles, providing the first principled, scenario-adaptive methodology for computational method selection in UTM fleet monitoring operations.
\end{abstract}

\begin{IEEEkeywords}
Unmanned aircraft systems, coordination detection, game theory, multi-objective optimization, Dynamic Mode Decomposition, Koopman operator, method selection, Pareto optimality, UTM
\end{IEEEkeywords}

\section{Introduction}
\label{sec:intro}

The rapid growth of commercial unmanned aerial vehicle (UAV) operations (including package delivery, infrastructure inspection, aerial survey, and urban air mobility) has created an urgent need for automated methods to detect coordinated fleet behavior among groups of aircraft observed in shared airspace.
An Unmanned Aircraft System Traffic Management (UTM) operator monitoring a service volume must answer three questions: (1)~Which aircraft are operating as \emph{coordinated fleets} versus flying independently? (2)~Which aircraft serve as the \emph{route-leads}---the navigation controllers whose decisions propagate to the rest of the fleet? (3)~What is the \emph{fastest reliable} way to answer (1) and (2)?

The first question is critical for airspace capacity management: a coordinated fleet of six drones occupies a single virtual traffic block, not six independent separation-managed slots.
The second question addresses safety assurance: if the route-lead experiences sensor degradation, GPS denial, or a lost command-and-control link, the entire fleet's trajectory integrity is compromised, requiring targeted contingency intervention.
The third question---method selection---is critical for operational deployment in real-time UTM systems processing thousands of position updates per second during peak traffic, yet it has received little attention in the literature.

Multiple detection algorithms exist (cross-recurrence analysis, Koopman operator methods, information-theoretic baselines, graph signal processing), each with distinct speed--accuracy characteristics that vary with fleet size, observation duration, and coordination structure.
No single method dominates across all operating conditions.

This paper addresses the method selection problem through three contributions:

\begin{enumerate}
    \item \textbf{End-to-end detection pipeline:} We implement and benchmark eight distinct coordination detection algorithms spanning four algorithmic families, establishing their stochastic performance profiles across 400+ randomized operating scenarios representative of diverse UTM traffic conditions.
    
    \item \textbf{Game-theoretic decision framework:} We formulate method selection as a two-player zero-sum stochastic game where the Monitor (choosing methods and parameters) plays against Nature (choosing the unknown traffic scenario). The minimax solution yields a \emph{robust mixed strategy}---a portfolio of methods that guarantees worst-case performance regardless of which traffic conditions materialize.
    
    \item \textbf{Multi-objective Pareto optimization:} We solve the joint optimization over coordination accuracy, route-lead identification accuracy, and computational speed using NSGA-II, producing a Pareto front of non-dominated method configurations from which operators select based on mission-specific priority weights.
\end{enumerate}

The key insight is that operational method selection is \emph{not} a single-objective problem.
Real-time deconfliction during peak delivery windows demands speed (favoring Graph Signal Processing at $43{,}061\times$ real-time), while post-incident forensic analysis prioritizes accuracy (favoring CRQA), and safety certification requires identifying the fleet's control node (favoring the Physics Baseline).
Our framework unifies these competing objectives into a single decision surface parameterized by operational priority weights $(w_{\text{coord}}, w_{\text{lead}}, w_{\text{speed}})$.

\subsection{Related Work}

Detection of coordinated behavior in multi-agent systems has been approached through dynamical systems theory \cite{schmid2010,budisic2012}, information-theoretic methods \cite{granger1969,schreiber2000}, recurrence analysis \cite{marwan2007}, and graph signal processing \cite{dong2016,ortega2018, kalofolias2016}.
However, prior work treats method evaluation as a benchmarking exercise---comparing methods post-hoc---rather than providing a \emph{prescriptive framework} for method selection under uncertainty.

Multi-objective optimization via evolutionary algorithms (NSGA-II \cite{deb2002}) has been applied to UAV path planning and resource allocation in air traffic systems, but not to the algorithm selection layer.
Game-theoretic formulations for robust decision-making under uncertainty exist in traffic management \cite{filar1996} and sensor resource allocation, but have not been applied to the computational pipeline design problem addressed here.

\subsection{Paper Organization}

Section~\ref{sec:formulation} formalizes the observation model and detection objectives.
Section~\ref{sec:methods} summarizes the eight candidate detection methods.
Section~\ref{sec:framework} presents the game-theoretic decision framework---the central contribution.
Section~\ref{sec:experiments} describes the experimental protocol.
Section~\ref{sec:results} demonstrates the framework's utility with empirical results.
Section~\ref{sec:conclusion} concludes.

\section{Problem Formulation}
\label{sec:formulation}

\subsection{Observation Model}

We consider $N$ UAVs observed over $T$ seconds at 1~Hz sampling rate via ADS-B, Remote ID broadcasts, or ground-based surveillance sensors.
Each aircraft $k$ has position $\mathbf{p}_k(t) \in \mathbb{R}^3$ at time $t$.
The trajectory is reduced to a scalar distance-to-reference time series:
\begin{equation}
    x_k(t) = \|\mathbf{p}_k(t) - \mathbf{p}_{\text{ref}}\|_2, \quad k = 1, \ldots, N
    \label{eq:observation}
\end{equation}
where $\mathbf{p}_{\text{ref}}$ is a fixed reference point (e.g., a vertiport, UTM service volume entry waypoint, or controlled airspace boundary) and $\|\cdot\|_2$ denotes the Euclidean norm.
This reduction preserves temporal dynamics (acceleration, waypoint transitions, corridor entry timing) while eliminating directional dependence.

\subsection{Three-Objective Detection Problem}

The coordination detection task has three competing objectives:
\begin{enumerate}
    \item \textbf{Coordination detection accuracy:} F1 score over all $\binom{N}{2}$ aircraft pairs:
    \begin{equation}
        f_1 = \text{F1}_{\text{coord}} = \frac{2 \cdot |\hat{\mathcal{C}} \cap \mathcal{C}|}{|\hat{\mathcal{C}}| + |\mathcal{C}|}
        \label{eq:coord_f1}
    \end{equation}
    where $\mathcal{C} \subseteq \{(i,j) : 1 \leq i < j \leq N\}$ is the set of truly coordinated aircraft pairs (members of the same fleet), $\hat{\mathcal{C}}$ is the set of pairs detected as coordinated by the algorithm, and $|\cdot|$ denotes set cardinality.
    
    \item \textbf{Route-lead identification accuracy:} F1 score over aircraft indices, quantifying the ability to identify the fleet's navigation controller---the aircraft whose trajectory decisions propagate to all fleet members:
    \begin{equation}
        f_2 = \text{F1}_{\text{lead}} = \frac{2 \cdot |\hat{\mathcal{L}} \cap \mathcal{L}|}{|\hat{\mathcal{L}}| + |\mathcal{L}|}
        \label{eq:leader_f1}
    \end{equation}
    where $\mathcal{L} \subseteq \{1, \ldots, N\}$ is the true set of route-lead aircraft and $\hat{\mathcal{L}}$ is the detected set.
    
    \item \textbf{Computational speed:} Real-time multiplier, measuring how many times faster than real-time the method executes:
    \begin{equation}
        f_3 = \text{xRT} = \frac{T}{\Delta t_{\text{compute}}}
        \label{eq:xrt}
    \end{equation}
    where $\Delta t_{\text{compute}}$ is the wall-clock time (in seconds) required to process the $T$-second observation window. Values $f_3 > 1$ indicate faster-than-real-time processing.
\end{enumerate}

The decision problem is: given an uncertain traffic scenario characterized by $(N, N_L, T, \text{coord\_pct})$---where $N_L$ is the number of route-lead aircraft (independent fleet coordinators) and $\text{coord\_pct} \in [0, 100]$ is the percentage of the observation window during which fleet coordination is active---and a set of candidate methods $\mathcal{M} = \{m_1, \ldots, m_8\}$ each with tunable parameters $\boldsymbol{\theta}_m \in \Theta_m$ (method-specific parameter spaces), select the method--parameter combination $(m^*, \boldsymbol{\theta}^*)$ that optimally balances $(f_1, f_2, f_3)$ according to operational priorities.

\subsection{Why Single-Method Selection Fails}

As will be demonstrated empirically in Section~\ref{sec:methods} and Table~\ref{tab:no_dominance}, no single method achieves the best performance on all three objectives simultaneously.
This non-dominance structure motivates a multi-objective approach: coordination detection accuracy, route-lead identification accuracy, and computational speed form a three-way trade-off where each method excels on at most two objectives.

\section{Candidate Detection Methods}
\label{sec:methods}

We evaluate eight methods spanning four algorithmic families.
Full mathematical details are provided in Appendix~\ref{app:methods}; here we summarize each method's key mechanism and computational complexity.
Table~\ref{tab:no_dominance} summarizes the mean empirical performance of all eight methods, confirming that no single method dominates across all objectives.

\begin{table}[t]
\centering
\caption{Non-Dominance of Detection Methods (400 Monte Carlo Trials, seed=333)}
\label{tab:no_dominance}
\begin{tabular}{lccc}
\toprule
\textbf{Method} & \textbf{Coord.\ F1} & \textbf{Lead.\ F1} & \textbf{Med.\ xRT} \\
\midrule
Physics Baseline & 0.551 & \textbf{0.422} & 59$\times$ \\
CRQA & \textbf{0.615} & 0.418 & 411$\times$ \\
Koopman/DMD & 0.208 & 0.255 & 1,378$\times$ \\
Streaming DMD & 0.208 & 0.282 & 1,383$\times$ \\
Koopman Phase & 0.208 & 0.284 & 1,392$\times$ \\
mrDMD & 0.212 & 0.276 & 1,384$\times$ \\
DMDc & 0.162 & 0.111 & 6,532$\times$ \\
Graph SP & 0.478 & 0.234 & \textbf{43,061}$\times$ \\
\bottomrule
\end{tabular}
\end{table}

\subsection{Pairwise Statistical Methods}

\subsubsection{Physics Baseline (DTW + Granger Causality + Transfer Entropy)}
\label{sec:physics}

Combines Dynamic Time Warping \cite{sakoe1978} for nonlinear alignment, Granger causality \cite{granger1969} for directed coupling, and Transfer Entropy \cite{schreiber2000} for model-free information flow.
\textbf{Complexity:} $O(N^2 T^2)$.
\textbf{Strength:} Gold-standard causal inference.
\textbf{Weakness:} Prohibitive for $N > 20$ in real-time.

\subsubsection{Cross-Recurrence Quantification Analysis (CRQA)}
\label{sec:crqa}

For each trajectory pair, constructs binary cross-recurrence matrices \cite{zbilut1998,marwan2007}:
\begin{equation}
    \mathbf{CR}_{ij}(m, n) = \Theta\left(\varepsilon - |x_i(m) - x_j(n)|\right)
    \label{eq:crp}
\end{equation}
where $\Theta(\cdot)$ is the Heaviside step function (0 if the argument is negative, 1 otherwise), $\varepsilon > 0$ is the recurrence threshold radius, and $x_i(m)$ and $x_j(n)$ are the scalar time series values of aircraft $i$ and $j$ at times $m$ and $n$ respectively.
Route-leads are detected by finding  UAVs with a sufficiently high diagonal recurrence rate with other aircraft. 
The time-delay embedding lag $\tau$ is selected automatically per trajectory using the mutual information function \cite{fraser1986,cao1997} ($O(NT\tau_{\max})$ with $\tau_{\max} = \min(50, T/4)$).
The embedding dimension $d$ is found via the false nearest neighbors (FNN) algorithm \cite{kennel1992} ($O(N\hat{M}^2 d^2)$, $\hat{M} = \min(T, 300)$). The recurrence threshold $\varepsilon$ is treated as a tunable parameter (Table~\ref{tab:sweep_params}).
\textbf{Complexity:} $O(N^2 T^2 + NT\tau_{\max} + N\hat{M}^2 d^2)$.
\textbf{Strength:} Best coordination detection F1 (0.615).
\textbf{Weakness:} Quadratic scaling; weak route-lead identification.

\subsection{Batch Operator-Theoretic Methods}

\subsubsection{Batch Dynamic Mode Decomposition (Koopman/DMD)}
\label{sec:koopman}
Performs global spectral decomposition \cite{schmid2010,tu2014} by first time-delay embedding scalar observables into delay-coordinate Koopman space \cite{arbabi2017,brunton2017chaos}.
It then uses a dense thin SVD followed by rank-$r$ truncation to compute $\tilde{\mathbf{A}}$, the reduced-order dynamics matrix whose eigenvalues approximate the Koopman operator spectrum:
\begin{equation}
    \tilde{\mathbf{A}} = \mathbf{U}_r^T \mathbf{X}_1 \mathbf{V}_r \boldsymbol{\Sigma}_r^{-1}.
    \label{eq:dmd}
\end{equation}
Here, $\mathbf{X}_1$ is the time-shifted data matrix while $\mathbf{U}_r$, $\boldsymbol{\Sigma}_r$, $\mathbf{V}_r$ are the rank-$r$ truncated SVD components of the unshifted data matrix $\mathbf{X}_0 \approx \mathbf{U}_r \boldsymbol{\Sigma}_r \mathbf{V}_r^T$.
Coordination is determined by comparing least-squares modal coefficient time series across aircraft pairs over a bounded range of temporal lags.
The delay embedding dimension $d$ is selected per aircraft via FNN, adding $O(N\hat{M}^2 d^2)$ preprocessing cost.
\textbf{Complexity:} $O(NTd^2 + NdTr + r^3 + N^2rTL_{\text{max}} + NTL_{\text{max}} + N\hat{M}^2 d^2)$, where $L_{\text{max}}$ is the largest coordination or leadership lag. The $NTd^2$ term accounts for the dense thin SVD used by the implementation; $N^2rTL_{\text{max}}$ accounts for the pairwise modal-sequence lag search.
\textbf{Key advantage:} Uses one shared spectral decomposition for all aircraft.

\subsubsection{Phase-Based Route-Lead Detection (Koopman Phase)}
\label{sec:phase}

Same DMD decomposition, but we've developed a different heuristic for leader detection. Here, the leader is detected by aggregating each UAV's phase advance across oscillatory modes \cite{mezic2005,budisic2012}:
\begin{equation}
    L_k = \sum_{m \in \mathcal{M}_{\text{osc}}} w_m \cdot \angle\left(e^{i(\theta_k^{(m)} - \bar{\theta}^{(m)})}\right)
    \label{eq:phase_leader}
\end{equation}
where $\mathcal{M}_{\text{osc}}$ is the set of oscillatory DMD modes (those satisfying $|\lambda_m| > 0.5$ and $|\arg(\lambda_m)| > 0.05$), $w_m$ is the energy weight of mode $m$, $\theta_k^{(m)}$ is the phase of aircraft $k$'s projection onto mode $m$, $\bar{\theta}^{(m)}$ is the mean phase across all aircraft, and $\angle(\cdot)$ extracts the argument of a complex number.

Aircraft with consistently positive $L_k$ (phase advance) are identified as route-leads.
As with Koopman/DMD, FNN selects $d$ per aircraft at cost $O(N\hat{M}^2 d^2)$.
\textbf{Complexity:} $O(NTd^2 + NdTr + r^3 + N^2rTL_{\text{max}} + Ndr + N\hat{M}^2 d^2)$. The $N^2rTL_{\text{max}}$ term is the modal-sequence coordination search; $Ndr$ accounts for the per-aircraft initial-condition phase calculations across the retained modes.
\textbf{Strength:} Modestly improves route-lead identification within the DMD family while preserving the coordination scores of batch and Streaming DMD.

\subsubsection{Multi-Resolution DMD (mrDMD)}
\label{sec:mrdmd}

Recursive DMD \cite{kutz2016} separating slow (transit trend) from fast (coordination) dynamics:
\begin{equation}
    \mathbf{X}_{\text{resid}} = \mathbf{X}_0 - \boldsymbol{\Phi}_{\text{slow}} \text{diag}(\mathbf{b}_{\text{slow}}) \mathbf{V}_{\text{slow}}
    \label{eq:mrdmd}
\end{equation}
where $\mathbf{X}_0$ is the original data matrix, $\boldsymbol{\Phi}_{\text{slow}}$ contains the DMD modes corresponding to slow dynamics (eigenvalues near $\lambda = 1$), $\mathbf{b}_{\text{slow}}$ are their amplitudes, and $\mathbf{V}_{\text{slow}}$ contains the corresponding Vandermonde time evolution. The residual $\mathbf{X}_{\text{resid}}$ isolates coordination-frequency dynamics for subsequent analysis.
FNN selects $d$ per aircraft at cost $O(N\hat{M}^2 d^2)$.
\textbf{Complexity:} $O(NTd^2 + NdT(r_0 + r_1) + N^2r_1TL_{\text{max}} + N\hat{M}^2 d^2)$ where $r_0, r_1$ are the ranks at successive resolution levels. The pairwise lagged-correlation coordination step uses the $r_1$ residual modes.
\textbf{Strength:} Offers multiscale decomposition with a modest coordination-detection advantage over other DMD variants.

\subsubsection{DMD with Control (DMDc)}
\label{sec:dmdc}

Models the route-lead as control input \cite{proctor2016}:
\begin{equation}
    \mathbf{x}_{t+1} = \mathbf{A}\mathbf{x}_t + \mathbf{B}\mathbf{u}_t
    \label{eq:dmdc}
\end{equation}
where $\mathbf{x}_t \in \mathbb{R}^{(N-N_L)d}$ is the delay-embedded state vector of the $N - N_L$ fleet members at time $t$, $\mathbf{u}_t \in \mathbb{R}^{N_L d}$ is the delay-embedded trajectory of the $N_L$ candidate route-leads (treated as exogenous inputs), $\mathbf{A}$ captures inter-member dynamics, and $\mathbf{B}$ quantifies the route-leads' influence on each fleet member. The use of Hankel-embedded state and control vectors (embedding dimension $d$ for both) is consistent with the Hankel-DMDc framework \cite{palma2025,brunton2022siam}.

We use the heuristic that the route-lead is the candidate $k$ maximizing $\|\mathbf{B}_k\|_F$, the Frobenius norm of the influence matrix when agent $k$ is hypothesized as route-lead.
FNN selects $d$ per aircraft at cost $O(N\hat{M}^2 d^2)$.
\textbf{Complexity:} $O(N^2 dTr + N\hat{M}^2 d^2)$ since all $N$ candidates are tested.
\textbf{Unique capability:} Per-member dependency mapping via $v_j = \|\mathbf{B}_{j\cdot}\|_F$, quantifying each aircraft's coupling to the route-lead's navigation decisions.

\subsection{Online Operator Methods}

\subsubsection{Streaming DMD}
\label{sec:streaming}

Reconstructs the DMD operator from incrementally updated Gram matrices \cite{hemati2014,zhang2019}:
\begin{equation}
    \mathbf{G}_{xx} \leftarrow \mathbf{G}_{xx} + x_t x_t^T
    \label{eq:streaming}
\end{equation}
where $\mathbf{G}_{xx} \in \mathbb{R}^{d \times d}$ is the accumulated outer-product matrix and $x_t \in \mathbb{R}^d$ is the $d$-dimensional delay-embedded state snapshot at time $t$. By constructing the DMD operator using accumulated Gram matrices, we avoid storing the full $d \times N(T-d)$ data matrix.
FNN selects $d$ per aircraft at cost $O(N\hat{M}^2 d^2)$ as a one-time preprocessing step before streaming begins.
\textbf{Complexity:} $O(NTd^2 + d^3 + r^3 + N^2rTL_{\text{max}} + Nd + N\hat{M}^2 d^2)$. The $NTd^2$ term is the Gram-matrix accumulation cost, while $N^2rTL_{\text{max}}$ accounts for batch post-processing of modal coefficient sequences for coordination; leader phase evaluation from the dominant mode costs $Nd$.
\textbf{Strength:} True online capability for continuous UTM monitoring.

\subsection{Graph-Topological Methods}

\subsubsection{Graph Signal Processing (Graph SP)}
\label{sec:graphsp}

Learns interaction topology \cite{dong2016,ortega2018, kalofolias2016} via convex optimization and identifies route-leads through Katz centrality \cite{katz1953}:
\begin{equation}
    \mathbf{c}_{\text{Katz}} = \left[(\mathbf{I} - \alpha_K \mathbf{W})^{-1} - \mathbf{I}\right]\mathbf{1}
    \label{eq:katz}
\end{equation}
where $\mathbf{I}$ is the $N \times N$ identity matrix, $\mathbf{W}$ is the learned adjacency (weight) matrix encoding pairwise interaction strengths, $\alpha_K < 1/\rho(\mathbf{W})$ is the attenuation factor (constrained below the spectral radius $\rho$ for convergence), and $\mathbf{1}$ is the all-ones vector. The $k$-th entry of $\mathbf{c}_{\text{Katz}}$ measures aircraft $k$'s influence across all path lengths in the learned fleet communication network.
\textbf{Complexity:} $O(N^2 Td + N^2 K + N^3)$ where $K$ is the number of proximal gradient iterations for topology learning.
\textbf{Unique capability:} Explicit fleet communication topology recovery.

\section{Game-Theoretic Decision Framework}
\label{sec:framework}

This section presents the central contribution: a principled framework for selecting among the eight candidate methods under scenario uncertainty, with provable worst-case guarantees.

\subsection{Decision Space}
\label{sec:decision_space}

Define the \emph{configuration space} $\mathcal{A}$ as the set of all feasible (method, parameter) combinations.
Each configuration is:
\begin{equation}
    a = (m, \boldsymbol{\theta}_m) \in \mathcal{A} = \bigcup_{m \in \mathcal{M}} \{m\} \times \Theta_m
    \label{eq:config_space}
\end{equation}
for some method $m \in \mathcal{M}$ with parameter vector $\boldsymbol{\theta}_m \in \Theta_m$.

The \emph{scenario space} $\mathcal{S}$ represents the unknown operating conditions:
\begin{equation}
    s = (N, N_L, \text{coord\_pct}, T) \in \mathcal{S}
    \label{eq:scenario_space}
\end{equation}
where $N$ is the number of aircraft in the monitoring volume, $N_L$ is the number of independent route-leads (fleet coordinators), $\text{coord\_pct}$ is the fraction of the observation time during which fleet coordination is active, and $T$ is the observation duration.

\subsection{Payoff Tensor}
\label{sec:payoff}

For each configuration--scenario pair $(a, s)$, the performance is a stochastic vector:
\begin{equation}
    \mathbf{f}(a, s) = \begin{bmatrix} f_1(a,s) \\ f_2(a,s) \\ f_3(a,s) \end{bmatrix} = \begin{bmatrix} \text{F1}_{\text{coord}} \\ \text{F1}_{\text{lead}} \\ \text{xRT} \end{bmatrix}
    \label{eq:payoff_vector}
\end{equation}

We estimate $\mathbb{E}[\mathbf{f}(a, s)]$ via Monte Carlo sampling (Section~\ref{sec:experiments}).
The payoff tensor $\mathcal{F} \in \mathbb{R}^{|\mathcal{A}| \times |\mathcal{S}| \times 3}$ stores these expected values.

\subsection{Scalarization}
\label{sec:scalarize}

Given operational priority weights $\mathbf{w} = (w_1, w_2, w_3)$ with $\sum_i w_i = 1$, we scalarize the payoff into a single utility:
\begin{equation}
    u(a, s; \mathbf{w}) = w_1 f_1(a,s) + w_2 f_2(a,s) + w_3 \cdot \tilde{f}_3(a,s)
    \label{eq:utility}
\end{equation}
where the normalized speed is:
\begin{equation}
    \tilde{f}_3 = \frac{\log(1 + \text{xRT}) - \log(1 + \text{xRT}_{\min})}{\log(1 + \text{xRT}_{\max}) - \log(1 + \text{xRT}_{\min})}
    \label{eq:norm_speed}
\end{equation}
The logarithmic normalization maps xRT (which spans $2\times$ to $934{,}718\times$) to $[0,1]$, preventing speed from dominating the utility at linear scale.

\subsection{Two-Player Zero-Sum Game Formulation}
\label{sec:game}

Let $\mathcal{A}_M$ be the set of feasible method configurations in $\mathcal{A}$ that the Monitor could choose, and let $\Delta(\cdot)$ denote the space of possible probability distributions over a given finite set.
We model the method selection problem as a finite two-player zero-sum game $\Gamma = (\mathcal{A}_M, \mathcal{S}, \mathbf{U})$:

\begin{itemize}
    \item \textbf{Player 1 (Monitor):} Selects a mixed strategy \\$\mathbf{p} \in \Delta(\mathcal{A}_M)$ over feasible method configurations.
    \item \textbf{Player 2 (Nature):} Selects a mixed strategy \\$\mathbf{q} \in \Delta(\mathcal{S})$ over scenarios, adversarially minimizing the Monitor's utility. Nature represents the uncertain traffic conditions---fleet sizes, coordination regimes, observation quality---that the UTM system cannot predict.
    \item \textbf{Payoff matrix:} $U_{ij} = u(a_i, s_j; \mathbf{w})$ for configuration $a_i$ and scenario $s_j$.
\end{itemize}

The Monitor seeks the \emph{maximin strategy}:
\begin{equation}
    \mathbf{p}^* = \arg\max_{\mathbf{p} \in \Delta(\mathcal{A}_M)} \min_{\mathbf{q} \in \Delta(\mathcal{S})} \mathbf{p}^\top \mathbf{U} \mathbf{q}
    \label{eq:maximin}
\end{equation}

By the minimax theorem \cite{vonneumann1928}, the game value $v^*~=~\mathbf{p}^{*\top} \mathbf{U}~\mathbf{q}^*$ exists and equals the guaranteed worst-case utility achievable by the Monitor's optimal mixed strategy, regardless of Nature's realization.

The solution $\mathbf{p}^*$ is a \emph{probability distribution over methods}---not a single method recommendation.
Operationally, this means:
\begin{itemize}
    \item In a single-shot decision, sample from $\mathbf{p}^*$ to select a method.
    \item With parallel compute budget, run the top-$k$ methods weighted by $\mathbf{p}^*$ and fuse outputs (for some user-chosen integer $k$).
    \item For continuous monitoring, alternate methods according to $\mathbf{p}^*$ across successive observation windows.
\end{itemize}

\subsection{Solution via Fictitious Play}
\label{sec:fictitious_play}

We solve the game using Brown's fictitious play algorithm \cite{brown1951}:
\begin{enumerate}
    \item Initialize empirical counts $\mathbf{n}_M(0) = \mathbf{e}_1$, $\mathbf{n}_N(0) = \mathbf{e}_1$. Here, $\mathbf{n}_M(t) \in \mathbb{R}^{|\mathcal{A}_M|}$ and $\mathbf{n}_N(t) \in \mathbb{R}^{|S|}$ are vectors that track how many times the Monitor and Nature respectively chose a particular scenario or configuration by time $t$.
    \item At iteration $t$:
    \begin{itemize}
        \item Monitor best-responds: \\$a^*(t) = \arg\max_i [\mathbf{U} \cdot \mathbf{n}_N(t\!-\!1)/t]_i$
        \item Nature best-responds: \\$s^*(t) = \arg\min_j [\mathbf{n}_M(t\!-\!1)^\top \mathbf{U}/t]_j$
        \item Update: $\mathbf{n}_M(t) \mathrel{+}= \mathbf{e}_{a^*(t)}$, similarly for Nature.
    \end{itemize}
    \item Converge when\\ $\left|\max_i \left[\mathbf{U} \frac{\mathbf{n}_N(t)}{t}\right]_i - \min_j \left[\frac{\mathbf{n}_M(t)^\top}{t} \mathbf{U}\right]_j\right| \leq \varepsilon$, i.e., the gap between the Monitor's best-response value against Nature's empirical play and Nature's best-response value against the Monitor's empirical play falls below tolerance $\varepsilon$. 
\end{enumerate}

Fictitious play converges to the minimax solution for finite games \cite{robinson1951} and is computationally efficient for our game dimensions ($\sim$8 methods $\times$ 244 observed scenarios).

\subsection{Scenario-Conditional Policy}
\label{sec:conditional}

When the scenario is \emph{known} (e.g., fleet size estimated from Remote ID broadcast counts), the game reduces to a single-column optimization:
\begin{equation}
    a^*(s) = \arg\max_{a \in \mathcal{A}_M} u(a, s; \mathbf{w})
    \label{eq:conditional}
\end{equation}
yielding a deterministic lookup table from traffic scenario features to optimal method.

\subsection{Multi-Objective Pareto Optimization}
\label{sec:pareto}

For operators who prefer the full trade-off surface, we identify the configurations in the \emph{Pareto front} via NSGA-II \cite{deb2002}:

\begin{definition}[Pareto Dominance]
Configuration $a_1$ dominates $a_2$ (written $a_1 \succ a_2$) if:
\begin{equation}
\forall i: \bar{f}_i(a_1) \geq \bar{f}_i(a_2) \;\; \text{and} \;\; \exists i: \bar{f}_i(a_1) > \bar{f}_i(a_2)
\label{eq:pareto}
\end{equation}
where $\bar{f}_i(a) = \mathbb{E}_s[f_i(a, s)]$ is the scenario-averaged performance. 
\end{definition}

The Pareto front $\mathcal{P} = \{a \in \mathcal{A} : \nexists a' \in \mathcal{A},\, a' \succ a\}$ is approximated using NSGA-II with population size 50, 80 generations, crossover (parameter blending within methods), and mutation (method switching at 10\%, parameter perturbation at 30\%). In the event where the resultant
configurations are not usable (e.g. when blending parameters results in a non-integer embedding dimension or rank), we round the parameters to the nearest usable value. 

\section{Experimental Design}
\label{sec:experiments}

\subsection{Trajectory Generation}

We employ a physically motivated route-lead/fleet-member model where $N_L$ route-lead aircraft navigate through 3--6 waypoints with cosine easing (Eq.~\ref{eq:cosine_easing_app} in Appendix~\ref{app:trajectory}).
Fleet members receive route updates via V2V datalink with communication and response delays $\delta_k \sim \mathcal{U}(2, 6)$~s, creating the characteristic temporal lag exploited by detection algorithms.
This delay budget reflects realistic V2V mesh latency (0.2--0.8~s), onboard path-planning recomputation (0.3--1.5~s), and flight controller response with airframe dynamics (1.0--2.5~s).

The coordination window spans the central \texttt{coord\_pct}\% of the observation period, representing the fleet in active coordinated transit; outside this window, aircraft fly independently along individual routes (e.g., last-mile delivery legs or return-to-hub segments), providing a null hypothesis for detection specificity.

\subsection{Monte Carlo Sensitivity Sweep}

To populate the payoff tensor $\mathcal{F}$, we execute $n = 400$ randomized trials (seed = 333, using CPU) drawing uniformly from the parameter ranges in Table~\ref{tab:sweep_params}.
Each trial records $(f_1, f_2, f_3)$ for all eight methods in both CPU and GPU modes, yielding $400 \times 8 = 3{,}200$ performance observations.

\begin{table}[h]
\centering
\caption{Sensitivity Sweep Parameter Ranges}
\label{tab:sweep_params}
\begin{tabular}{lll}
\toprule
\textbf{Category} & \textbf{Parameter} & \textbf{Range} \\
\midrule
\multirow{4}{*}{Scenario} & $N$ (aircraft) & $\{5, 10, 15, 20, 30, 50\}$ \\
 & $N_L$ (route-leads) & $\{1, 2, 3, 5\}$ \\
 & coord\_pct & $\{25, 50, 75, 100\}\%$ \\
 & $T$ (duration) & $\{50, 100, 200, 500\}$ s \\
\midrule
\multirow{4}{*}{Method} & $\varepsilon$ (CRQA) & $\{2, 5, 10, 15, 20, 30\}$ \\
 & $r$ (DMD rank) & $\{5, 10, 15, 20, 30\}$ \\
 & $d$ (delay embed) & $\{3, 5, 8, 10, 15\}$ \\
 & $n_{\text{levels}}$ (mrDMD) & $\{2, 3, 4\}$ \\
 & $\alpha$ (Graph SP) & $\{0.1, 0.5, 1.0, 2.0, 5.0\}$ \\
 & $\beta$ (Graph SP) & $\{0.01, 0.05, 0.1, 0.5, 1.0\}$ \\
\midrule
Baseline & $L$ (Granger lag) & $\{3, 5, 8, 10, 15\}$ \\
\bottomrule
\end{tabular}
\end{table}

\subsection{Sensitivity Quantification}

Parameter sensitivity is quantified via eta-squared ($\eta^2$), an effect-size measure from analysis of variance (ANOVA) that captures the proportion of total variance in a performance metric attributable to a given parameter:
\begin{equation}
    \eta^2 = \frac{SS_{\text{between}}}{SS_{\text{total}}} = \frac{\sum_{g=1}^{G} n_g (\bar{y}_g - \bar{y})^2}{\sum_{i=1}^{n} (y_i - \bar{y})^2}
    \label{eq:eta_sq}
\end{equation}
where $G$ is the number of distinct levels (values) the parameter takes, $n_g$ is the number of trials at level $g$, $\bar{y}_g$ is the mean performance metric (e.g., coordination F1) for trials at level $g$, $\bar{y}$ is the grand mean across all $n$ trials, and $y_i$ is the individual trial outcome.
The numerator $SS_{\text{between}}$ measures how much the group means deviate from the grand mean (variance explained by the parameter), while the denominator $SS_{\text{total}}$ measures total variance.
Values of $\eta^2 > 0.15$ indicate high sensitivity (the parameter explains more than 15\% of performance variance), $0.08 < \eta^2 \leq 0.15$ indicates moderate sensitivity, and $\eta^2 \leq 0.08$ indicates the parameter has minimal practical impact on that metric.
This quantification is computed independently for each (method, parameter, metric) triple, yielding the sensitivity heatmaps in Section~\ref{sec:results}.

\subsection{Decision Framework Protocol}

Given the populated payoff tensor, the decision framework:
\begin{enumerate}
    \item Constructs the method-level payoff matrix $\mathbf{U}$ (best per-method performance per scenario, aggregated over parameter settings)
    \item Solves the two-player game via 5,000 iterations of fictitious play
    \item Computes scenario-conditional recommendations for all 244 observed scenarios
    \item Runs NSGA-II (population size of 50; 80 generations) for Pareto front approximation
\end{enumerate}

We evaluate four operational priority profiles that span representative UTM use cases.
Each profile is defined by a weight triplet $\mathbf{w} = (w_{\text{coord}}, w_{\text{lead}}, w_{\text{speed}})$ where $w_{\text{coord}}$ is the importance of coordination detection accuracy (F1$_{\text{coord}}$), $w_{\text{lead}}$ is the importance of route-lead identification accuracy (F1$_{\text{lead}}$), and $w_{\text{speed}}$ is the importance of computational speed (normalized xRT). The weights satisfy $w_{\text{coord}} + w_{\text{lead}} + w_{\text{speed}} = 1$.
The four profiles are:
\begin{itemize}
    \item \textbf{Balanced} $(0.40, 0.35, 0.25)$: General-purpose UTM monitoring where all three objectives matter, with slight preference for fleet membership detection.
    \item \textbf{Speed-Priority} $(0.15, 0.15, 0.70)$: Peak delivery window operations where the method must execute fast enough for real-time corridor slot allocation, tolerating reduced accuracy.
    \item \textbf{Accuracy-Priority} $(0.50, 0.40, 0.10)$: Post-incident forensic analysis where detection fidelity is paramount (e.g., determining fleet membership after a near-miss event) and compute time is unconstrained.
    \item \textbf{Route-Lead Focus} $(0.20, 0.60, 0.20)$: Safety certification and contingency monitoring where identifying the fleet's navigation controller---the single point of failure for the formation---outweighs coordination detection.
\end{itemize}

\section{Results and Analysis}
\label{sec:results}

\subsection{Computational Scaling}

\begin{figure}[t]
\centering
\includegraphics[width=\columnwidth]{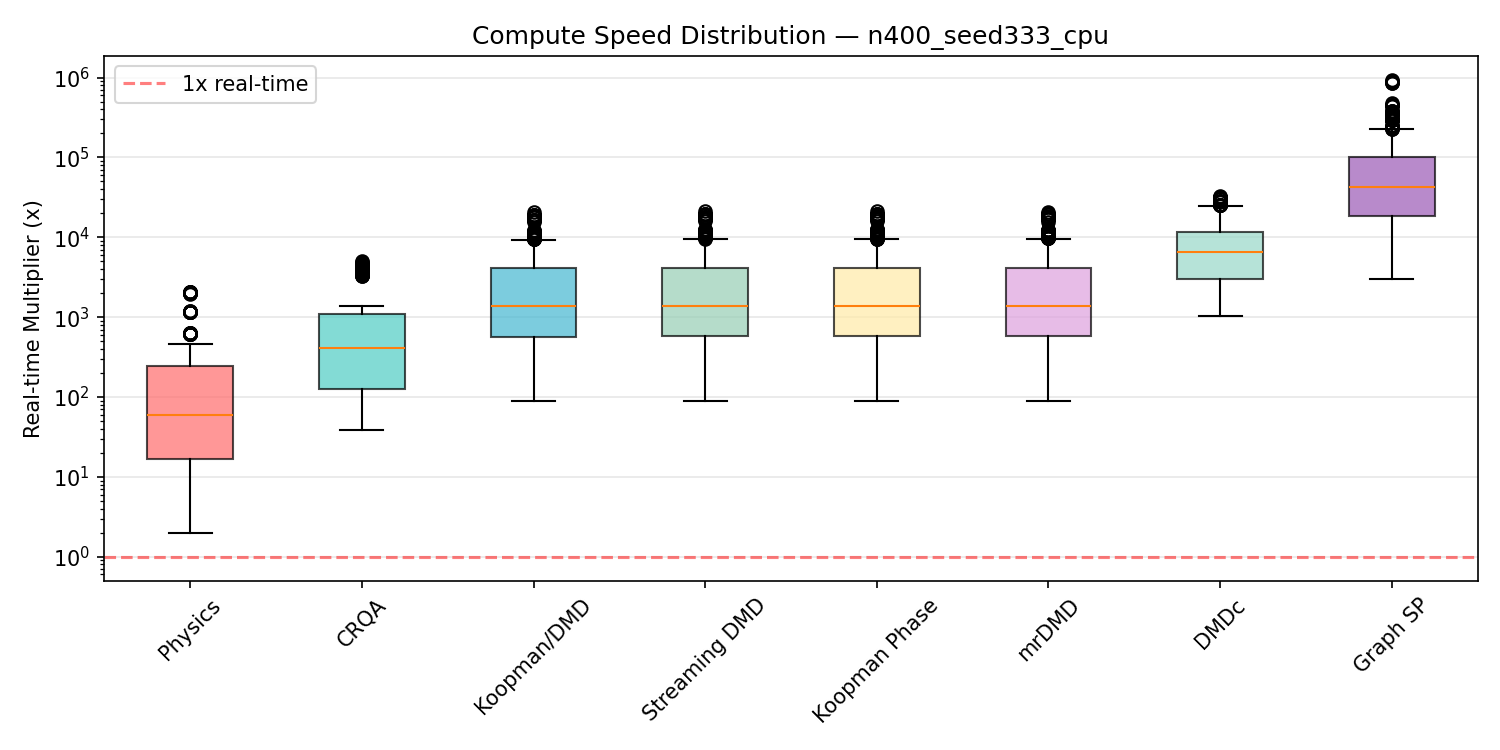}
\caption{Real-time multiplier distribution (log scale) across 400 Monte Carlo trials. All methods exceed real-time, but the gap between $O(N^2)$ and $O(N)$ methods spans two orders of magnitude.}
\label{fig:scaling}
\end{figure}

Figure~\ref{fig:scaling} shows the real-time multiplier distribution.
The observed CPU throughput differs substantially across method families:
\begin{itemize}
    \item $O(N^2 T^2)$ methods (Physics, CRQA): median xRT 59$\times$ and 411$\times$, respectively; at $N=50$, their medians fall to 5.1$\times$ and 45.0$\times$.
    \item Methods with an $O(N^2rTL_{\max})$ pairwise coordination stage (Koopman, mrDMD, Phase): median xRT 1,378--1,392$\times$; at $N=50$, their medians are 253--254$\times$.
    \item Graph-topological (Graph SP): median xRT 43,061$\times$; at $N=50$, the median is 9,710$\times$ and the observed minimum exceeds 2,966$\times$.
\end{itemize}

\subsection{Detection Accuracy Decoupling}

\begin{figure}[t]
\centering
\includegraphics[width=\columnwidth]{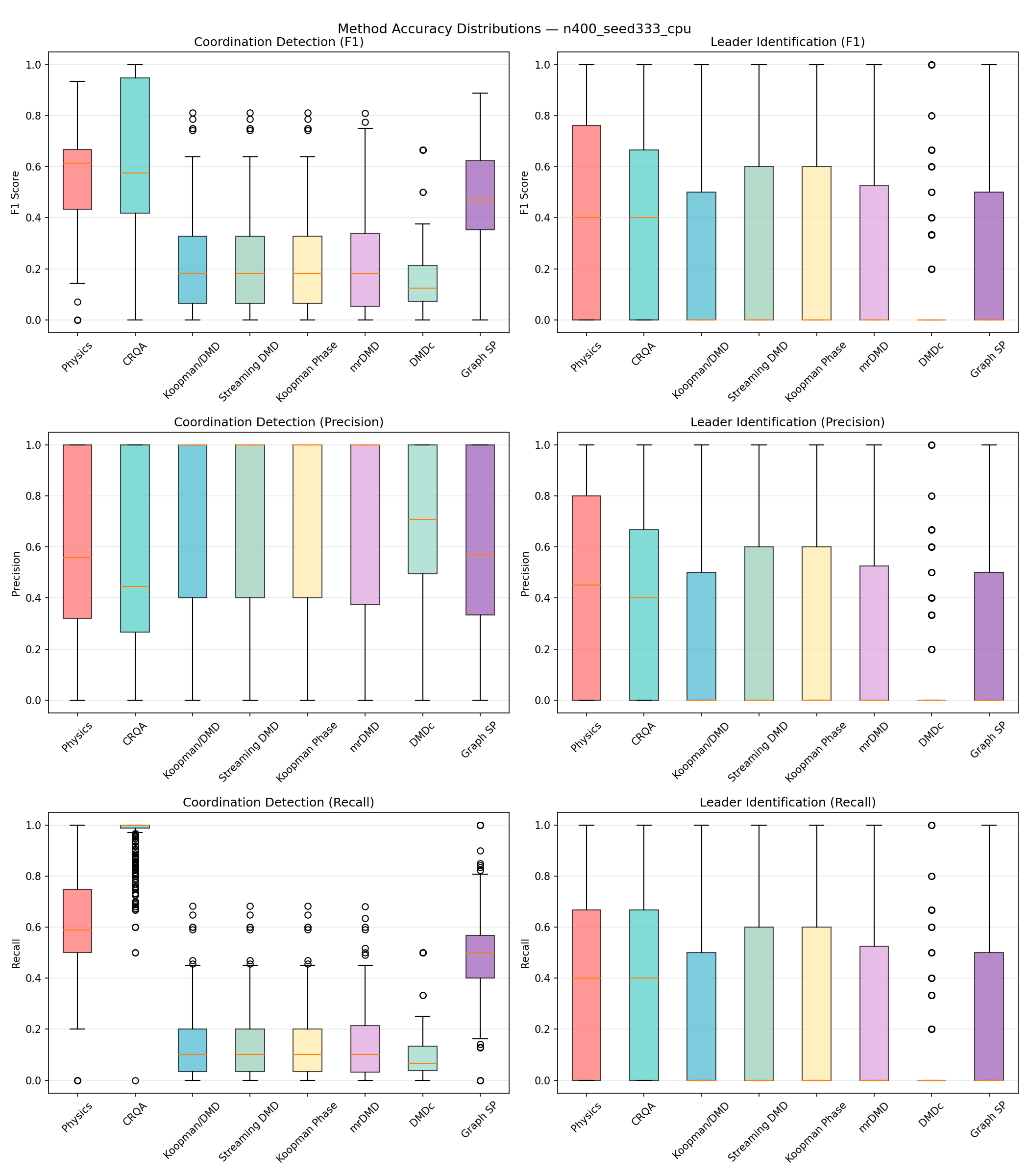}
\caption{Detection accuracy distributions. Left: Coordination F1, precision, and recall. Right: Route-lead F1, precision, and recall (identification of the fleet navigation controller). The decoupling between tasks motivates multi-objective optimization.}
\label{fig:accuracy}
\end{figure}

Figure~\ref{fig:accuracy} reveals that coordination detection and route-lead identification are \emph{decoupled tasks}: while CRQA has the best coordination F1 (0.615), it is instead the Physics Baseline that has the best route-lead F1 score (0.422).
Furthermore, three methods (Koopman/DMD, Streaming DMD, and Koopman Phase) have the same coordination F1 (.208), but different route-lead F1 scores (ranging from 0.255 to 0.284).
This within-family decoupling is what motivates the multi-objective framework, since optimizing for one objective does not automatically improve the other.

\subsection{Game-Theoretic Method Selection}
\label{sec:game_results}

Table~\ref{tab:game_results} presents the central result: the minimax mixed strategy for four operational priority profiles.

\begin{table*}[t]
\centering
\caption{Game-Theoretic Optimal Method Portfolios Under Different Operational Priorities}
\label{tab:game_results}
\begin{tabular}{lccc|ccl}
\toprule
\textbf{Profile} & $w_{\text{c}}$ & $w_{\text{l}}$ & $w_{\text{s}}$ & \textbf{Game Value} & \textbf{Primary (\%)} & \textbf{Supporting Methods} \\
\midrule
Balanced & 0.40 & 0.35 & 0.25 & 0.300 & Graph SP (59.9\%) &  Physics (33.1\%), CRQA (6.9\%) \\
Speed-Priority & 0.15 & 0.15 & 0.70 & 0.454 & Graph SP (100\%) \\
Accuracy-Priority & 0.50 & 0.40 & 0.10 & 0.278 & Graph SP (58.5\%) & Physics (32.7\%), CRQA (8.9\%) \\
Route-Lead Focus & 0.20 & 0.60 & 0.20 & 0.198 & Graph SP (65.4\%) & CRQA (22.6\%), Physics (11.7\%) \\
\bottomrule
\end{tabular}
\end{table*}


\textbf{Finding 1: Graph SP dominates all priority profiles.}
At 43,061$\times$ median speed, Graph SP offers the best trade-off across all priority settings, despite it's fairly average route-lead and 
coordination detection performance.

\textbf{Finding 2: Only the speed-priority recommends a single method at 100\%.}
Most game solutions are a \emph{mixture}, confirming that method diversity provides robustness against scenario uncertainty.
The game values range from 0.198 (Route-Lead Focus, hardest objective) to 0.454 (Speed-Priority).

\textbf{Finding 3: Physics Baseline appears in most profiles, and is usually favored over CRQA.}
Despite being the slowest method (59$\times$ median), Physics achieves 33.1\% allocation in the balanced profile, a 32.7\% allocation
in the accuracy profile, and 11.7\% in the route-lead-focused profile. Meanwhile, CRQA only garners 6.9\% of the balanced profile and 8.9\% of the accuracy profile.
These results are somewhat surprising since, according to Table \ref{tab:no_dominance}, CRQA performs better than the Physics Baseline when it comes to coordination 
detection accuracy and speed, and has a very similar route-lead F1 score. Thus, one would expect CRQA to be favored over the Physics Baseline in these profiles. To explain this discrepency, recall
that the maximin mixed strategy solves the stochastic game in which the Monitor responds to the \emph{worst-case} scenario created by Nature.
Meanwhile, the Monte Carlo simulations used to generate Table \ref{tab:no_dominance} used completely random scenarios. Thus, it's possible that the Physics
Baseline is more robust than CRQA against Nature's worst-case scenario (as the mixed strategy selects for) for the balanced and accuracy profiles, whereas CRQA performs
better in the \emph{average} scenario (represented by the Monte Carlo trials) for these profiles. 

\subsection{Challenging Scenario Analysis}

For the balanced profile $(0.40, 0.35, 0.25)$, Nature's approximate adversarial strategy after 5,000 fictitious-play iterations concentrates 98.42\% of its weight on three observed scenarios:
\begin{itemize}
    \item $N=50$, $N_L=5$, coord\_pct $=25\%$, $T=200$~s: 43.82\% of Nature's weight.
    \item $N=50$, $N_L=2$, coord\_pct $=25\%$, $T=50$~s: 36.78\%.
    \item $N=50$, $N_L=2$, coord\_pct $=25\%$, $T=500$~s: 17.82\%.
\end{itemize}

The common features are the largest tested fleet size and a short coordination fraction, rather than a single route-lead count or observation duration. This concentration is consistent with limited coordination evidence and the computational burden of large fleets.

\subsection{Scenario-Conditional Recommendations}

When the traffic scenario is known a priori (Table~\ref{tab:conditional}), the framework provides deterministic method selection. For the balanced weights $(0.40, 0.35, 0.25)$, the updated 400-trial CPU sweep yields 244 unique scenarios. The characteristic regimes summarize observed tendencies among the selected scenarios, not exclusive selection rules.

\begin{table}[t]
\centering
\caption{Scenario-Conditional Method Recommendations for Balanced Profile}
\label{tab:conditional}
\begin{tabular}{lcp{4cm}}
\toprule
\textbf{Method} & \textbf{\# Scenarios} & \textbf{Characteristic Regime} \\
\midrule
CRQA & 92 (37.7\%) & All fleet sizes; mostly few route-leads \\
Graph SP & 67 (27.5\%) & Mostly many route-leads ($N_L \geq 3$ in 49/67); often long observations ($T=500$~s in 30/67) \\
Physics & 46 (18.9\%) & Mostly medium--large fleets ($N~\geq~20$ in 29/46); short observations ($T \leq 200$~s in 43/46) \\
Koopman & 12 (4.9\%) & Mostly short coordination windows (25--50\% in 10/12); $T \geq 100$~s \\
K.~Phase & 9 (3.7\%) & Short observations ($T \leq 200$~s); multiple route-leads ($N_L \geq 2$); 50\% coordination in 5/9 \\
mrDMD & 6 (2.5\%) & Selected~fleets~($N~\in~\{5,10,15,30\}$); $T \leq 200$~s and 25--75\% coordination \\
DMDc & 6 (2.5\%) & Many route-leads ($N_L \geq 3$); selected fleets ($N \in \{5,15,30\}$) \\
Str.~DMD & 6 (2.5\%) & Short coordination windows (25--50\%); $T \leq 200$~s; $N~\in~\{10,15,30,50\}$ \\
\bottomrule
\end{tabular}
\end{table}


\subsection{Pareto Front}

The NSGA-II optimization identifies Pareto-optimal configurations, a representative sample of which is shown in Table~\ref{tab:pareto}.
The full Pareto front is given in the appendix.
The front is dominated by CRQA (accuracy end) and Graph SP (speed end), quantifying the exact accuracy
cost per unit of speed gain: switching from CRQA ($f_1 = 0.705$, xRT =834$\times$) to Graph SP
($f_1 = 0.469$, xRT = 80,659$\times$) yields a $96.71\times$ speed gain at the expense of 33.5\% coordination accuracy 
and 57.9\% route-lead accuracy.

\begin{table}[t]
\centering
\caption{Representative Pareto-Optimal Configurations for Balanced Profile}
\label{tab:pareto}
\begin{tabular}{llccc}
\toprule
\textbf{Method} & \textbf{Params} & $f_1$ & $f_2$ & \textbf{xRT} \\
\midrule

CRQA    & $\varepsilon\!=\!5.0$        & 0.705 & 0.561 & 834$\times$ \\
CRQA    & $\varepsilon\!=\!10.0$       & 0.567 & 0.617 & 1,200$\times$ \\
Graph~SP&                              & 0.469 & 0.242 & 80,659$\times$ \\
mrDMD   & $r\!=\!15, d\!=\!5, l\!=\!2$ & 0.336 & 0.417 & 2,927$\times$ \\
mrDMD   & $r\!=\!7, d\!=\!8, l\!=\!2$  & 0.233 & 0.425 & 2,793$\times$ \\
K~Phase & $r\!=\!11, d\!=\!8$          & 0.207 & 0.295 & 5,014$\times$ \\
Str.~DMD & $r\!=\!12, d\!=\!8$         & 0.200 & 0.309 & 4,654$\times$ \\
mrDMD   & $r\!=\!13, d\!=\!8, l\!=\!2$ & 0.151 & 0.480 & 3,766$\times$ \\
mrDMD   & $r\!=\!15, d\!=\!8, l\!=\!3$ & 0.142 & 0.358 & 5,633$\times$ \\
mrDMD   & $r\!=\!10, d\!=\!8, l\!=\!2$ & 0.132 & 0.520 & 4,557$\times$ \\
\bottomrule
\end{tabular}
\end{table}

\subsection{Sensitivity Analysis}

\begin{figure*}[t]
\centering
\includegraphics[width=\textwidth]{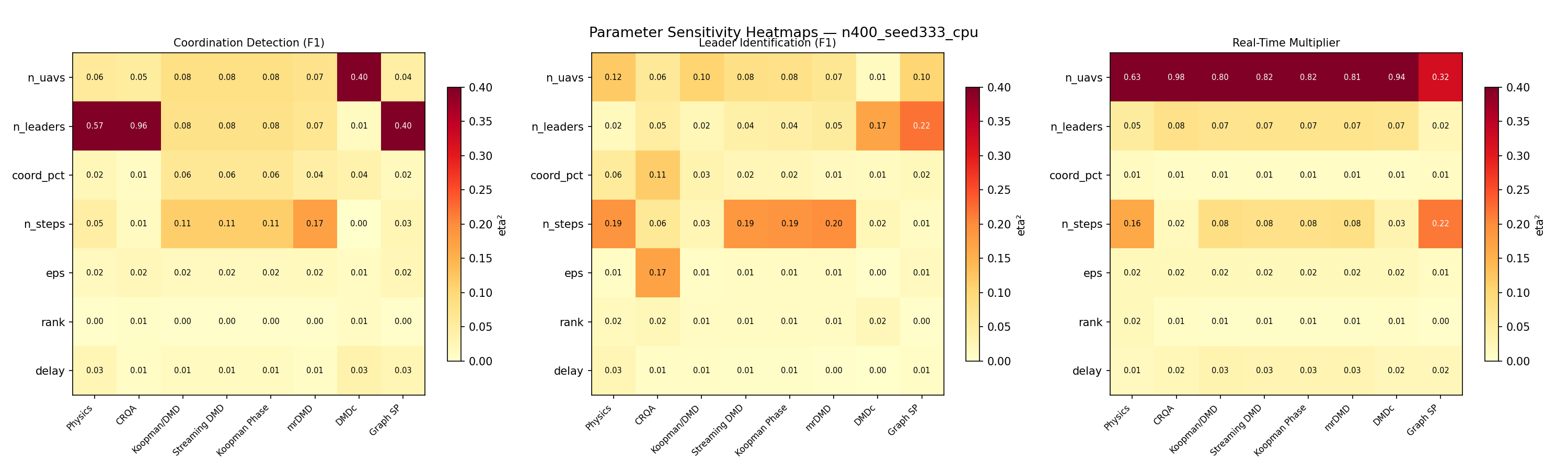}
\caption{Parameter sensitivity ($\eta^2$). Left: Coordination F1. Center: Route-lead F1. Right: Speed. Dark cells indicate strong parameter--performance coupling.}
\label{fig:sensitivity}
\end{figure*}

Figure~\ref{fig:sensitivity} shows eta-squared sensitivity heatmaps.
We see that the number of leaders, $N_L$, greatly influences coordination F1 ($\eta^2 = 0.40$--$0.97$) for the Physics Baseline, CRQA, and Graph Signal Processing. Meanwhile fleet size, $N$, critically affects DMDc's coordination F1, and dominates speed for all methods ($\eta^2 = 0.32$--$0.98$). The number of steps also affects speed for Graph SP ($\eta^2 = 0.22$).

\section{Discussion}
\label{sec:discussion}

\subsection{Framework Generality}

The decision framework is method-agnostic: it operates on the payoff tensor without assuming specific algorithm internals.
New detection methods are incorporated by evaluating them across the scenario grid and adding rows to $\mathcal{F}$.
New coordination models (distributed consensus, relay leadership, dynamic role-switching) extend the column space.
The game solution adapts instantaneously to new weight vectors without re-running detection algorithms.

\subsection{Sequential Deployment Strategy}

For the specific use case of monitoring coordinated UAV fleet operations, the CPU sweep recommends a staged approach:
\begin{enumerate}
    \item \textbf{Stage 1---Rapid screening} (speed-critical): Deploy Graph SP for preliminary fleet coordination screening (approximately 0.008--0.031~s across 66 trials with 50 aircraft and 50--500~s observation windows). Treat positive detections as candidates for confirmation rather than definitive capacity counts.
    \item \textbf{Stage 2---Confirmation and route-lead identification}: When accuracy is prioritized, use CRQA to confirm coordination (mean coordination F1 of 0.615 across the sweep). Then (if supported by the scenario-conditional policy) select Koopman Phase or Streaming DMD for route-lead identification.
    \item \textbf{Stage 3---Dependency mapping}: Once route-lead confirmed, deploy DMDc for exploratory per-member coupling estimates.
    \item \textbf{Stage 4---Adaptation}: If traffic conditions shift (new fleet enters volume, weather degrades, fleet begins dispersing), re-query the conditional lookup table.
\end{enumerate}

\subsection{Operational Implications for UTM}

The framework addresses several practical UTM challenges:
\begin{itemize}
    \item \textbf{Capacity management:} Accurate fleet detection enables correct capacity counting---a 6-drone coordinated fleet occupies one virtual traffic block rather than six independent separation-managed slots.
    \item \textbf{Contingency handling:} Identifying the route-lead enables targeted contingency commands. If the route-lead enters a degraded state (GPS denied, lost C2 link), the UTM system can issue a fleet-wide hold rather than treating each aircraft independently.
    \item \textbf{Compliance verification:} Operators are required to declare fleet coordination in flight plans; the detection framework provides independent verification from surveillance data when declarations are absent or inconsistent.
\end{itemize}

\subsection{Computational Cost of the Framework}

The decision framework itself (game solution + NSGA-II) executes in approximately 3.3 seconds on the experimental platform: an NVIDIA DGX Spark\footnote{Specifications per NVIDIA DGX Spark Hardware Documentation, \url{https://docs.nvidia.com/dgx/dgx-spark/hardware.html}, accessed May 2026.} powered by the NVIDIA GB10 Grace Blackwell Superchip.
The system features a 20-core ARM CPU (10$\times$ Cortex-X925 + 10$\times$ Cortex-A725), 128~GB LPDDR5x unified memory (273~GB/s bandwidth, 256-bit interface), and an integrated NVIDIA Blackwell GPU with 6,144 CUDA cores and 5th-generation Tensor Cores (CUDA~13.0, Driver~580.159).
The expensive step---populating the payoff tensor (400 configurations $\times$ 8 methods $\times$ 2 modes = 6,400 evaluations)---required approximately 2.7 hours utilizing 4 parallel CPU workers on the same platform.
Once the payoff tensor is computed (a one-time offline cost), the game solution adapts to new weight vectors instantaneously without re-running detection algorithms.

\subsection{Limitations}

The current framework assumes stationary traffic scenarios within each observation window.
Extension to sequential games (where the method can be switched mid-observation as traffic evolves) and to evasion-aware coordination models (where fleet operators deliberately mask coordination signatures to circumvent airspace capacity restrictions) are natural next steps.
The payoff tensor quality depends on the Monte Carlo sample size and parameter grid resolution; sparse coverage in high-dimensional parameter spaces may miss performance pockets.
Additionally, the current trajectory model assumes a single coordination archetype (route-lead/fleet-member); real UTM environments may exhibit more complex patterns including dynamic role-switching, hierarchical sub-fleets, and heterogeneous aircraft with differing response characteristics.

\section{Conclusion}
\label{sec:conclusion}

We have presented a game-theoretic decision framework for selecting coordination detection methods in multi-UAV fleet monitoring scenarios.
The framework resolves the fundamental speed--accuracy trade-off through:

\begin{enumerate}
    \item A \emph{stochastic game formulation} providing guaranteed worst-case performance via minimax mixed strategies, ensuring reliable monitoring regardless of uncertain traffic conditions.
    
    \item \emph{Empirical validation} demonstrating that optimal portfolios shift with operational priorities: Graph SP (100\%) for speed-focus, but only (59.9\%) for balanced monitoring.
    \item A \emph{Pareto front} quantifying the accuracy--speed trade-off: $96.71\times$ speed gain at 33.5\% coordination accuracy cost along the CRQA-to-Graph~SP transition.
    
    \item A \emph{reusable infrastructure} adapting instantaneously to new priority weights without re-running detection algorithms---enabling dynamic reconfiguration as UTM operational tempo changes.
\end{enumerate}

The framework provides the first principled, scenario-adaptive methodology for computational method selection in UTM fleet monitoring operations, applicable beyond the route-lead/fleet-member use case to any setting where multiple detection algorithms compete under uncertain operating conditions.

\bibliographystyle{IEEEtran}

\appendices

\section{Detailed Mathematical Derivations}
\label{app:methods}

This appendix provides the complete mathematical details for each of the eight detection methods summarized in Section~\ref{sec:methods}.

\subsection{Physics Baseline (DTW + Granger Causality + Transfer Entropy)}

The Physics Baseline combines three complementary statistical measures:

\textbf{Dynamic Time Warping (DTW):} For trajectories $x_i(t)$ and $x_j(t)$, DTW finds the optimal warping path $W~=~\{w_1, \ldots, w_K\}$ minimizing:
\begin{equation}
    \text{DTW}(x_i, x_j) = \min_W \sum_{k=1}^K d(w_k)
    \label{eq:dtw_app}
\end{equation}
where $d(w_k)$ is the Euclidean distance between the pair of time indices aligned at warping step $k$. The minimization is solved via dynamic programming in $O(T^2)$.

\textbf{Granger Causality:} Tests whether past values of $x_i$ improve prediction of $x_j$ beyond $x_j$'s own history:
\begin{equation}
    x_j(t) = \sum_{\ell=1}^L a_\ell \, x_j(t-\ell) + \sum_{\ell=1}^L b_\ell \, x_i(t-\ell) + \varepsilon(t)
    \label{eq:granger_app}
\end{equation}
where $L$ is the maximum lag order, $a_\ell$ are autoregressive coefficients, $b_\ell$ are cross-regressive coefficients, and $\varepsilon(t)$ is white noise. If the $b_\ell$ coefficients are jointly significant (F-test, $p < 0.05$), aircraft $i$ Granger-causes aircraft $j$.

\textbf{Transfer Entropy:} Information-theoretic measure of directed information flow:
\begin{equation}
    T_{i \to j} = \sum p(x_j^{t+1}, x_j^t, x_i^t) \log \frac{p(x_j^{t+1} | x_j^t, x_i^t)}{p(x_j^{t+1} | x_j^t)}
    \label{eq:te_app}
\end{equation}
where $p(\cdot)$ denotes probability distributions estimated via histogram binning. The summation is over all observed value combinations.

\subsection{Cross-Recurrence Quantification Analysis (CRQA)}

According to \cite{marwan2007}, the diagonal recurrence rate profile:

\begin{equation}
    \text{RR}(\tau) = \frac{1}{N-\tau} \sum_{i = 1}^{N- \tau} \boldsymbol{R}_{i, i + \tau},
\end{equation}
quantifies synchronization at each lag $\tau$. Here,
\begin{equation}
    \mathbf{R}_{i, j}(\varepsilon) = \Theta\left(\varepsilon - \|x_i - x_j\|\right) 
\end{equation}
for some threshold distance $\varepsilon$.
The lag $\tau^*$ maximizing $\text{RR}(\tau)$ indicates the temporal offset between aircraft; if $\tau^* > 0$, aircraft $i$ leads aircraft $j$.

We create a heuristic for leadership scores by accumulating weighted directional evidence:
\begin{equation}
    L_k = \sum_{\substack{(i,j): \\ \text{conf}_{ij} > \theta}} w_{ij} \cdot \indicator [\text{aircraft } k \text{ leads in pair } (i,j)]
    \label{eq:crqa_leader_app}
\end{equation}
where $w_{ij} = \text{RR}(\tau^*_{ij}) \cdot \text{conf}_{ij}$, and the confidence is $\text{conf}_{ij}~=~\text{RR}(\tau^*_{ij})~-~\text{median}_\tau(\text{RR}(\tau))$.

\subsection{Batch Dynamic Mode Decomposition (Koopman/DMD)}

The global data matrix is formed by horizontally concatenating scaled versions of the delay-embedded Hankel matrices~\cite{arbabi2017,brunton2017chaos} of all $N$ aircraft:
\begin{equation}
    \mathbf{X} = \begin{bmatrix} \alpha_1 H(x_1) & \alpha_2 H(x_2) & \cdots & \alpha_N H(x_N) \end{bmatrix} \in \mathbb{R}^{d \times N(T-d+1)}
    \label{eq:hankel_app}
\end{equation}
where $H(x_k) \in \mathbb{R}^{d \times (T-d+1)}$ is the Hankel (delay-embedding) matrix of aircraft $k$ with embedding dimension $d$, and $\alpha_k$ is the ratio of $H_(x_k)$'s last column's magnitude and $H_(x_1)$'s last column's magnitude. Each column of $\mathbf{X}$ is a $d$-dimensional delay-embedded snapshot; columns from different aircraft are concatenated along the time axis.

DMD finds the best-fit linear operator satisfying $\mathbf{X}_1~\approx~\mathbf{A}~\mathbf{X}_0$, where $\mathbf{X}_0$ and $\mathbf{X}_1$ are time-shifted submatrices of $\mathbf{X}$. The DMD modes are:
\begin{equation}
    \boldsymbol{\Phi} = \mathbf{U}_r \mathbf{W}
    \label{eq:dmd_modes_app}
\end{equation}
where $\mathbf{W}$ contains the eigenvectors of $\tilde{\mathbf{A}} = \mathbf{U}_r^T \mathbf{X}_1 \mathbf{V}_r \boldsymbol{\Sigma}_r^{-1}$.

Since the Koopman modes $\boldsymbol{\Phi} \in \mathbb{C}^{d \times r}$ live in the shared delay-embedding space, we obtain each aircraft's modal coefficient sequence using a least-squares fit of its scaled delay snapshots to the shared modes:
\begin{equation}
    \mathbf{C}_k = \arg\min_{\mathbf{C}} \|\boldsymbol{\Phi}\mathbf{C} - \alpha_k H(x_k)\|_F^2.
    \label{eq:modal_coeff_app}
\end{equation}
Here $H(x_k) \in \mathbb{R}^{d \times T'}$ is aircraft $k$'s Hankel matrix, $T'~=~T~-~d$, and column $t$ of $\mathbf{C}_k$ contains its least-squares coefficients for the DMD modes at snapshot $t$. For each aircraft pair, the method computes centered, normalized cross-correlations between corresponding modal coefficient sequences over lags $|\ell|\leq L_{\max}$, retaining one representative from each complex-conjugate mode pair. The correlation between aircraft $i$ and $j$ with respect to mode $m$ is:
\begin{equation}
    \rho_{ij,m}(\ell) = \frac{\left|\sum_{t\in I_\ell} \overline{\tilde{c}_{i,m}^{(\ell)}(t)}\,\tilde{c}_{j,m}^{(\ell)}(t+\ell)\right|}{\left(\sum_{t\in I_\ell}|\tilde{c}_{i,m}^{(\ell)}(t)|^2\sum_{t\in I_\ell}|\tilde{c}_{j,m}^{(\ell)}(t+\ell)|^2\right)^{1/2}}.
    \label{eq:modal_corr_app}
\end{equation}
Here, $I_\ell$ is the valid overlap after shifting by lag $\ell$, and $\tilde{c}_{k,m}^{(\ell)}$ denotes coefficient sequence for aircraft $k$ with respect to mode $m$ centered over that overlap. Modal correlations are weighted by the geometric mean of the two aircraft's modal energies; the maximum weighted score over the tested lags is compared with a threshold to classify the pair as coordinated. Thus, coordination is based on temporal similarity and relative lag of the modal coefficient sequences.

\subsection{Streaming DMD}

Per-aircraft Hankel matrices are horizontally concatenated as in Eq.~\ref{eq:hankel_app}, giving snapshot matrices $\mathbf{X}_0, \mathbf{X}_1 \in \mathbb{R}^{d \times N(T-d)}$ with state dimension $d$ (not $Nd$); each column is a single $d$-dimensional delay-embedded snapshot. The Gram matrices $\mathbf{G}_{xx}, \mathbf{G}_{xy} \in \mathbb{R}^{d \times d}$ therefore scale with the embedding dimension only, independent of $N$.

Initialization uses the first $r+1$ snapshots to form seed Gram matrices:
\begin{equation}
    \mathbf{G}_{xx}^{(0)} = \mathbf{X}_{\text{init}} \mathbf{X}_{\text{init}}^T, \quad \mathbf{G}_{xy}^{(0)} = \mathbf{Y}_{\text{init}} \mathbf{X}_{\text{init}}^T
    \label{eq:streaming_init_app}
\end{equation}
where $\mathbf{X}_{\text{init}}$ and $\mathbf{Y}_{\text{init}}$ are the unshifted and shifted snapshot matrices from the warm-up window.

At each subsequent time step, rank-1 updates maintain the statistics:
\begin{equation}
    \mathbf{G}_{xx} \leftarrow \mathbf{G}_{xx} + x_t x_t^T, \quad \mathbf{G}_{xy} \leftarrow \mathbf{G}_{xy} + y_t x_t^T
    \label{eq:streaming_update_app}
\end{equation}

The dynamics operator is recovered as:
\begin{equation}
    \hat{\mathbf{A}} = \mathbf{G}_{xy} \mathbf{G}_{xx}^{\dagger}
    \label{eq:streaming_operator_app}
\end{equation}
where $(\cdot)^\dagger$ denotes the Moore--Penrose pseudoinverse, computed via SVD of $\mathbf{G}_{xx}$ truncated to rank $r$.
In practice we compute the rank-$r$ Galerkin projection \cite{tu2014} directly from the SVD of $\mathbf{G}_{xx} = \mathbf{U}_g \mathbf{S}_g \mathbf{U}_g^T$:
\begin{equation}
    \tilde{\mathbf{A}} = \mathbf{U}_{g,r}^T \mathbf{G}_{xy} \mathbf{U}_{g,r} \mathbf{S}_{g,r}^{-1}
    \label{eq:streaming_reduced_app}
\end{equation}
where $\mathbf{U}_{g,r} \in \mathbb{R}^{d \times r}$ are the leading $r$ eigenvectors of $\mathbf{G}_{xx}$ and $\mathbf{S}_{g,r}$ contains the corresponding eigenvalues. This follows from substituting $\mathbf{G}_{xx}^{\dagger} \approx \mathbf{U}_{g,r} \mathbf{S}_{g,r}^{-1} \mathbf{U}_{g,r}^T$ into $\hat{\mathbf{A}}$ and applying $\mathbf{U}_{g,r}^T \mathbf{U}_{g,r} = \mathbf{I}_r$. The DMD modes are then $\boldsymbol{\Phi} = \mathbf{U}_{g,r} \mathbf{W}$, where $\mathbf{W}$ contains the eigenvectors of $\tilde{\mathbf{A}}$.

\textit{Batch Extraction Complexity Note:} The online accumulation of the Gram matrices costs $O(NTd^2)$, followed by an $O(d^3)$ SVD of the $d\times d$ Gram matrix and an $O(r^3)$ reduced eigendecomposition. Batch coordination post-processing first fits modal coefficient sequences by least squares, then checks each aircraft pair over up to $L_{\max}$ lags; its dominant pairwise cost is $O(N^2rTL_{\max})$. FNN dimension selection adds $O(N\hat{M}^2d^2)$.

\subsection{Koopman Phase (Phase-Based Route-Lead Detection)}

The DMD decomposition is identical to the batch method: per-aircraft Hankel matrices are horizontally concatenated (Eq.~\ref{eq:hankel_app}), yielding $\mathbf{X}_0 \in \mathbb{R}^{d \times N(T-d)}$, modes $\boldsymbol{\Phi} \in \mathbb{R}^{d \times r}$, and eigenvalues $\{\lambda_m\}$ via the same SVD and reduced operator $\tilde{\mathbf{A}}$.


We select oscillatory modes as those whose eigenvalue satisfies $|\lambda_m| > 0.5$ and $|\arg(\lambda_m)| > 0.05$. The theoretical ideal is $|\lambda_m| \approx 1$ (neutrally stable, non-decaying modes on the unit circle \cite{mezic2005}), but DMD eigenvalues computed from short, noisy trajectories are biased away from the unit circle \cite{hemati2017}. We therefore use the threshold $|\lambda_m| > 0.5$ as a practical relaxation that retains modes with meaningful oscillatory energy while excluding heavily damped transients.



The phase of aircraft $k$ in mode $m$ is extracted via its initial-condition projection onto the mode:
\begin{equation}
    \theta_k^{(m)} = \arg\left(\boldsymbol{\phi}_m \cdot \mathbf{h}_k(0)\right)
    \label{eq:agent_phase_app}
\end{equation}
where $\boldsymbol{\phi}_m \in \mathbb{R}^d$ is the $m$-th column of the mode matrix $\boldsymbol{\Phi}$ and $\mathbf{h}_k(0) \in \mathbb{R}^d$ is the initial delay-embedded snapshot of aircraft $k$. The initial-condition projection preserves phase information.

To see why, note that the DMD reconstruction \cite{tu2014} approximates the state as a superposition of modes $\mathbf{h}(t) \approx \sum_j b_j \boldsymbol{\phi}_j \lambda_j^t$; projecting onto mode $m$ (and invoking the Koopman eigenfunction evolution property $U^t g_m = \lambda_m^t g_m$ \cite{budisic2012,mezic2005}) isolates mode $m$'s contribution to aircraft $k$'s trajectory:

\begin{equation}
    \boldsymbol{\phi}_m^T \mathbf{h}_k(t) \approx z_k^{(m)} \lambda_m^t = z_k^{(m)} |\lambda_m|^t e^{i\angle(\lambda_m)t}
    \label{eq:phase_evolution_app}
\end{equation}
where $z_k^{(m)} = \boldsymbol{\phi}_m \cdot \mathbf{h}_k(0)$ is the initial-condition projection. Taking the argument of both sides:
\begin{equation}
    \arg\!\left(\boldsymbol{\phi}_m^T \mathbf{h}_k(t)\right) = \underbrace{\arg(z_k^{(m)})}_{\theta_k^{(m)}} + \angle(\lambda_m)\,t
    \label{eq:phase_argument_app}
\end{equation}
The second term $\angle(\lambda_m)\,t$ is a common phase drift shared by \emph{all} aircraft in mode $m$. Consequently, the relative phases $\theta_k^{(m)} - \theta_j^{(m)}$ are time-invariant: the ordering of aircraft by phase does not depend on which snapshot time $t$ is used. Thus, the initial condition $t = 0$ is sufficient to determine each aircraft's phase offset within the mode's oscillatory cycle.

The group mean phase (circular mean) for mode $m$ is:
\begin{equation}
    \bar{\theta}^{(m)} = \arg\left(\sum_{k=1}^N e^{i\theta_k^{(m)}}\right)
    \label{eq:mean_phase_app}
\end{equation}

Aircraft with consistently positive phase lead $L_k > 0$ (Eq.~\ref{eq:phase_leader}) are ahead in the oscillatory coordination cycle and are identified as route-leads.

\subsection{Multi-Resolution DMD (mrDMD)}

Both resolution levels apply horizontal Hankel stacking (Eq.~\ref{eq:hankel_app}): the Level-0 and Level-1 data matrices are each of shape $d \times N(T-d)$, with modes $\boldsymbol{\Phi} \in \mathbb{R}^{d \times r}$.

At Level~0, standard DMD identifies slow modes with eigenvalues satisfying $|\angle \lambda| < \delta$ (low frequency threshold). The slow-dynamics reconstruction is:
\begin{equation}
    \mathbf{X}_{\text{slow}} = \boldsymbol{\Phi}_{\text{slow}} \, \text{diag}(\mathbf{b}_{\text{slow}}) \, \mathbf{V}_{\text{slow}}
    \label{eq:slow_recon_app}
\end{equation}
where $\boldsymbol{\Phi}_{\text{slow}}$ are the slow DMD modes, $\mathbf{b}_{\text{slow}}$ are their initial amplitudes (computed via least-squares), and $\mathbf{V}_{\text{slow}}~\in~\mathbb{C}^{r_{\text{slow}} \times N(T-d)}$ is the Vandermonde matrix with column $j$ equal to $\lambda^j$.

At Level~1, DMD is applied to the residual $\mathbf{X}_{\text{resid}}~=~\mathbf{X}_0~-~\mathbf{X}_{\text{slow}}$, extracting coordination-frequency dynamics free from transit trend contamination.

\subsection{DMD with Control (DMDc)}

Unlike the other Koopman methods, DMDc uses \emph{vertical} stacking: the delay-embedded Hankel matrices of the $N-N_L$ fleet members are stacked row-wise to form $\mathbf{X}_0~\in~\mathbb{R}^{(N-N_L)d \times (T-d)}$, and those of the $N_L$ route-leads form $\mathbf{U}_0 \in \mathbb{R}^{N_L d \times (T-d)}$. This is required by the DMDc formulation, which must identify $\mathbf{A}$ and $\mathbf{B}$ separately from a single augmented regression. The resulting structure is equivalent to the Hankel-DMDc framework of \cite{palma2025,brunton2022siam} with equal state and input delay depth $d-1$ and scalar (1-D) per-aircraft observables.

The augmented data matrix stacks state and control:
\begin{equation}
    \boldsymbol{\Omega} = \begin{bmatrix} \mathbf{X}_0 \\ \mathbf{U}_0 \end{bmatrix} \in \mathbb{R}^{((N-N_L)d + N_L d) \times (T-d)}
    \label{eq:dmdc_stack_app}
\end{equation}
where $\mathbf{X}_0 \in \mathbb{R}^{(N-N_L)d \times (T-d)}$ contains the delay-embedded trajectories of the $N - N_L$ fleet members and $\mathbf{U}_0~\in~\mathbb{R}^{N_L d \times (T-d)}$ contains the delay-embedded trajectories of the $N_L$ hypothesized route-leads.

The system matrices are recovered jointly \cite{proctor2016}:
\begin{equation}
    [\mathbf{A} \mid \mathbf{B}] = \mathbf{X}_1 \boldsymbol{\Omega}^{\dagger} = \mathbf{X}_1 \mathbf{V}_r \boldsymbol{\Sigma}_r^{-1} \mathbf{U}_r^T
    \label{eq:dmdc_solve_app}
\end{equation}
via truncated SVD of $\boldsymbol{\Omega}$, expanding the pseudoinverse as $\boldsymbol{\Omega}^\dagger~=~\mathbf{V}_r \boldsymbol{\Sigma}_r^{-1}~\mathbf{U}_r^T$ (cf.\ Eq.~(28) of \cite{proctor2016}).

Per-member dependency scores quantify coupling strength:
\begin{equation}
    v_j = \|\mathbf{B}_{j\cdot}\|_F = \sqrt{\sum_{\ell=1}^d B_{j,\ell}^2}
    \label{eq:dependency_app}
\end{equation}
High-$v_j$ aircraft are most tightly coupled to the route-lead's navigation decisions.

\subsection{Graph Signal Processing (Graph SP)}

\textbf{Step 1: Pairwise Distance Matrix.} For normalized trajectory vectors $\mathbf{z}_i, \mathbf{z}_j \in \mathbb{R}^{T}$ of aircraft $i$ and $j$, we form the mean squared distance matrix:
\begin{equation}
    Z_{ij} = \frac{1}{T}\|\mathbf{z}_i - \mathbf{z}_j\|^2, \quad Z_{ii} := 0
    \label{eq:gsp_sim_app}
\end{equation}
The matrix $\mathbf{Z} \in \mathbb{R}^{N\times N}$ enters the graph learning objective directly as the smoothness penalty weight: large $Z_{ij}$ discourages placing a strong edge between dissimilar aircraft.

\textbf{Step 2: Graph Learning.} Following \cite{kalofolias2016}, we learn the sparse adjacency matrix $\mathbf{W}$ by minimizing:
\begin{equation}
    \min_{\mathbf{W} \geq 0,\, \mathrm{diag}(\mathbf{W})=0} \; \mathbf{1}^\top(\mathbf{W} \circ \mathbf{Z})\mathbf{1} - \alpha\,\mathbf{1}^\top\log(\mathbf{W}\mathbf{1}) + \frac{\beta}{2}\|\mathbf{W}\|_F^2
    \label{eq:gsp_obj_app}
\end{equation}
The first term penalizes strong edges between dissimilar aircraft; the log-barrier $-\alpha\mathbf{1}^\top\log(\mathbf{W}\mathbf{1})$ prevents isolated nodes by driving all degrees positive ($\alpha > 0$); and the Frobenius term ($\beta > 0$) ensures uniqueness. We minimize via projected gradient descent with step size $\eta > 0$:
\begin{equation}
    \mathbf{W}^{(k+1)} = \Pi_+\!\left[\mathbf{W}^{(k)} - \eta\,\nabla f(\mathbf{W}^{(k)})\right]
    \label{eq:gsp_prox_app}
\end{equation}
where $\nabla f(\mathbf{W}) = \mathbf{Z} - \alpha\bigl(\mathbf{d}^{-1}\mathbf{1}^\top + \mathbf{1}(\mathbf{d}^{-1})^\top\bigr) + \beta\mathbf{W}$, $\mathbf{d} = \mathbf{W}\mathbf{1}$ is the degree vector, $\mathbf{d}^{-1}$ denotes the element-wise inverse, and $\Pi_+[\cdot] = \max(\cdot, 0)$ with diagonal zeroed enforces the constraints after each step.

\textbf{Step 3: Directional Leadership.} For each significant edge $(i,j)$ where $W_{ij}$ is larger than some threshold $\theta_W \in \mathbb{R}$, temporal lead-lag is assessed via one-step-shifted correlations:
\begin{equation}
    \rho^+_{ij} = \text{corr}(\mathbf{z}_i[1{:}], \mathbf{z}_j[{:}{-}1]), \quad \rho^-_{ij} = \text{corr}(\mathbf{z}_i[{:}{-}1], \mathbf{z}_j[1{:}])
    \label{eq:gsp_lag_app}
\end{equation}
If $\rho^+_{ij} > \rho^-_{ij}$, aircraft $i$ temporally leads aircraft $j$. The directional out-strength and in-strength are:
\begin{equation}
    S^{\text{out}}_i = \sum_{j: \rho^+_{ij} > \rho^-_{ij}} W_{ij}, \quad
    S^{\text{in}}_i  = \sum_{j: \rho^+_{ji} > \rho^-_{ji}} W_{ji}
    \label{eq:gsp_outstrength_app}
\end{equation}

\textbf{Final Score:} A convex combination of structural centrality and net temporal outflow, each min-max normalized:
\begin{equation}
    \ell_i = 0.6 \cdot \frac{c_{\text{Katz},i} - \min_k c_{\text{Katz},k}}{\max_k c_{\text{Katz},k} - \min_k c_{\text{Katz},k}} + 0.4 \cdot \frac{\Delta S_i - \min_k \Delta S_k}{\max_k \Delta S_k - \min_k \Delta S_k}
    \label{eq:gsp_score_app}
\end{equation}
where $\Delta S_i = S^{\text{out}}_i - S^{\text{in}}_i$ is the net temporal outflow (aircraft that lead more neighbors than they follow have $\Delta S_i > 0$).

\section{Trajectory Generation Model}
\label{app:trajectory}

Route-lead aircraft navigate through $W = 3$--6 randomly generated waypoints $\{\mathbf{p}^{(w)}\}_{w=1}^W$ in 3D space, representing corridor entry points, altitude transitions, and delivery zone boundaries. Between consecutive waypoints, motion follows a cosine easing interpolation:
\begin{equation}
    \mathbf{p}_{\text{lead}}(t) = \mathbf{p}^{(w)} + \frac{1}{2}\left(1 - \cos\left(\pi \cdot \frac{t - t_w}{t_{w+1} - t_w}\right)\right) \cdot \left(\mathbf{p}^{(w+1)} - \mathbf{p}^{(w)}\right)
    \label{eq:cosine_easing_app}
\end{equation}
for $t \in [t_w, t_{w+1}]$, where $t_w$ and $t_{w+1}$ are the arrival times at waypoints $w$ and $w+1$. This produces smooth acceleration/deceleration profiles characteristic of real multirotor flight paths, with zero velocity at waypoints.

Fleet members receive route updates via V2V datalink after stochastic communication and response delays $\delta_k \sim \mathcal{U}(2, 6)$~s, creating the temporal lag structure exploited by detection algorithms. Damped, zero-mean Gaussian noise is added to all positions to simulate GPS and navigation uncertainty.

The coordination window spans the central $\texttt{coord\_pct}\%$ of the $T$-second observation; outside this window, aircraft fly independently along individual routes (last-mile delivery legs, return-to-hub segments), providing a null hypothesis for detection specificity.

\section{Full Pareto Front}

Table~\ref{tab:full_front} shows all the configurations in the first Pareto front for the balanced priority profile.
\begin{table}[t]
\centering
\caption{Full Pareto Front for Balanced Profile}
\label{tab:full_front}
\begin{tabular}{llccc}
\toprule
\textbf{Method} & \textbf{Params} & $f_1$ & $f_2$ & \textbf{xRT} \\
\midrule

CRQA    & $\varepsilon\!=\!5.0$        & 0.705 & 0.561 & 834$\times$ \\
CRQA    & $\varepsilon\!=\!5.3$        & 0.688 & 0.558 & 842$\times$ \\
CRQA    & $\varepsilon\!=\!5.7$        & 0.672 & 0.557 & 857$\times$ \\
CRQA    & $\varepsilon\!=\!20.0$       & 0.648 & 0.338 & 1,387$\times$ \\
CRQA    & $\varepsilon\!=\!19.2$       & 0.634 & 0.363 & 1,311$\times$ \\
CRQA    & $\varepsilon\!=\!18.2$       & 0.621 & 0.383 & 1,234$\times$ \\
CRQA    & $\varepsilon\!=\!8.3$        & 0.610 & 0.581 & 1,016$\times$ \\
CRQA    & $\varepsilon\!=\!16.6$       & 0.606 & 0.397 & 1,122$\times$ \\
CRQA    & $\varepsilon\!=\!13.9$       & 0.596 & 0.415 & 1,072$\times$ \\
CRQA    & $\varepsilon\!=\!13.3$       & 0.595 & 0.439 & 1,103$\times$ \\
CRQA    & $\varepsilon\!=\!12.7$       & 0.592 & 0.466 & 1,128$\times$ \\
CRQA    & $\varepsilon\!=\!10.5$       & 0.581 & 0.589 & 1,150$\times$ \\
CRQA    & $\varepsilon\!=\!9.6$        & 0.579 & 0.598 & 1,160$\times$ \\
CRQA    & $\varepsilon\!=\!9.8$        & 0.575 & 0.605 & 1,174$\times$ \\
CRQA    & $\varepsilon\!=\!10.0$       & 0.567 & 0.617 & 1,200$\times$ \\
Graph~SP&                              & 0.469 & 0.242 & 80,659$\times$ \\
mrDMD   & $r\!=\!15, d\!=\!5, l\!=\!2$ & 0.336 & 0.417 & 2,927$\times$ \\
mrDMD   & $r\!=\!15, d\!=\!6, l\!=\!2$ & 0.294 & 0.373 & 3,764$\times$ \\
mrDMD   & $r\!=\!13, d\!=\!4, l\!=\!2$ & 0.292 & 0.244 & 4,277$\times$ \\
mrDMD   & $r\!=\!6, d\!=\!8, l\!=\!2$  & 0.260 & 0.427 & 2,380$\times$ \\
mrDMD   & $r\!=\!6, d\!=\!3, l\!=\!3$  & 0.242 & 0.309 & 4,582$\times$ \\
mrDMD   & $r\!=\!7, d\!=\!8, l\!=\!2$  & 0.233 & 0.425 & 2,793$\times$ \\
K~Phase & $r\!=\!8, d\!=\!8$           & 0.221 & 0.315 & 3,802$\times$ \\
Str.~DMD & $r\!=\!10, d\!=\!8$         & 0.211 & 0.283 & 5,543$\times$ \\
mrDMD   & $r\!=\!14, d\!=\!7, l\!=\!2$ & 0.209 & 0.497 & 3,502$\times$ \\
K~Phase & $r\!=\!11, d\!=\!8$          & 0.207 & 0.295 & 5,014$\times$ \\
mrDMD   & $r\!=\!8, d\!=\!8, l\!=\!2$  & 0.203 & 0.449 & 3,719$\times$ \\
mrDMD   & $r\!=\!13, d\!=\!7, l\!=\!2$ & 0.203 & 0.498 & 3,615$\times$ \\
Str.~DMD & $r\!=\!12, d\!=\!8$         & 0.200 & 0.309 & 4,654$\times$ \\
mrDMD   & $r\!=\!13, d\!=\!7, l\!=\!3$ & 0.175 & 0.335 & 4,751$\times$ \\
mrDMD   & $r\!=\!15, d\!=\!8, l\!=\!2$ & 0.171 & 0.533 & 3,320$\times$ \\
mrDMD   & $r\!=\!14, d\!=\!7, l\!=\!3$ & 0.166 & 0.353 & 4,735$\times$ \\
mrDMD   & $r\!=\!13, d\!=\!8, l\!=\!2$ & 0.151 & 0.480 & 3,766$\times$ \\
mrDMD   & $r\!=\!15, d\!=\!8, l\!=\!3$ & 0.142 & 0.358 & 5,633$\times$ \\
mrDMD   & $r\!=\!10, d\!=\!8, l\!=\!2$ & 0.132 & 0.520 & 4,557$\times$ \\
\bottomrule
\end{tabular}
\end{table}

\section{Complexity Summary}
\label{app:complexity}

Table~\ref{tab:complexity_app} summarizes computational complexity for all methods.

\begin{table*}[h]
\centering
\caption{Computational Complexity of Detection Methods}
\label{tab:complexity_app}
\begin{tabular}{llll}
\toprule
\textbf{Method} & \textbf{Time Complexity} & \textbf{Route-Lead Detection} & \textbf{Key Advantage} \\
\midrule
Physics (DTW+GC+TE) & $O(N^2 T^2)$ & Causal inference & Gold-standard accuracy \\
CRQA & $O(N^2 T^2 + NT\tau_{\max} + N\hat{M}^2 d^2)$ & Lag peak in RR profile & Robust to noise \\
Koopman/DMD & $O(NTd^2 + NdTr + r^3 + N^2rTL + NTL + N\hat{M}^2 d^2)$ & Centroid lag-correlation & Global decomposition \\
Streaming DMD & $O(NTd^2 + d^3 + r^3 + N^2rTL + Nd + N\hat{M}^2 d^2)$ & Dominant-mode phase & Online processing \\
Koopman Phase & $O(NTd^2 + NdTr + r^3 + N^2rTL + Ndr + N\hat{M}^2 d^2)$ & Modal phase & Phase-based leadership \\
mrDMD & $O(NTd^2 + NdT(r_0+r_1) + N^2r_1TL + N\hat{M}^2 d^2)$ & Residual modal phase & Multi-timescale isolation \\
DMDc & $O(N^2 dTr + N\hat{M}^2 d^2)$ & Control influence norm & Dependency mapping \\
Graph SP & $O(N^2 Td + N^2 K + N^3)$ & Katz centrality + lag & Topological inference \\
\bottomrule
\end{tabular}
\end{table*}

\noindent Where $N$ = number of aircraft, $T$ = trajectory length (samples), $d$ = delay embedding dimension (FNN-selected), $r$~=~DMD rank, $r_0$ and $r_1$ are the ranks at successive mrDMD levels, $L$ = maximum lag in modal-sequence coordination or centroid leadership, $K$ = proximal gradient iterations, $\tau_{\max} = \min(50, T/4)$ = maximum lag for mutual information delay selection, and $\hat{M} = \min(T, M_s)$ with $M_s~=~300$~=~subsampled trajectory length for FNN computation. The $N^2rTL$ terms account for evaluating up to $r$ modal correlations over $T$ samples and $L$ lags for every aircraft pair. The dense thin SVD used by the batch Koopman methods and at each mrDMD level costs $O(NTd^2)$; truncated operator projection costs $O(NdTr)$. For Streaming DMD, $O(NTd^2)$ is Gram-matrix accumulation, while $O(d^3)$ is the Gram-matrix SVD.

\begin{IEEEbiography}[{\includegraphics[width=1in,height=1.25in,clip,keepaspectratio]{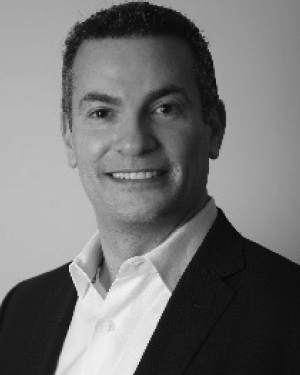}}]{Christian G. Manasseh}
received the M.Eng.\ degree in information technology from the Massachusetts Institute of Technology, Cambridge, MA, USA, in 1999, and the Ph.D.\ degree in systems engineering from the University of California at Berkeley, Berkeley, CA, USA, in 2010. He is currently the Founder of Mobius Logic Inc., Tysons, VA, USA, a company specializing in artificial intelligence and data science research and development. He has coauthored multiple research papers on dynamic time warping, game-theoretic decision frameworks, agent-based modeling, and artificial intelligence for human decision-making. His research interests include operator-theoretic methods for dynamical systems, multi-objective optimization under uncertainty, and GPU-accelerated scientific computing for autonomous systems.
\end{IEEEbiography}

\begin{IEEEbiography}[{\includegraphics[width=1in,height=1.25in,clip,keepaspectratio]{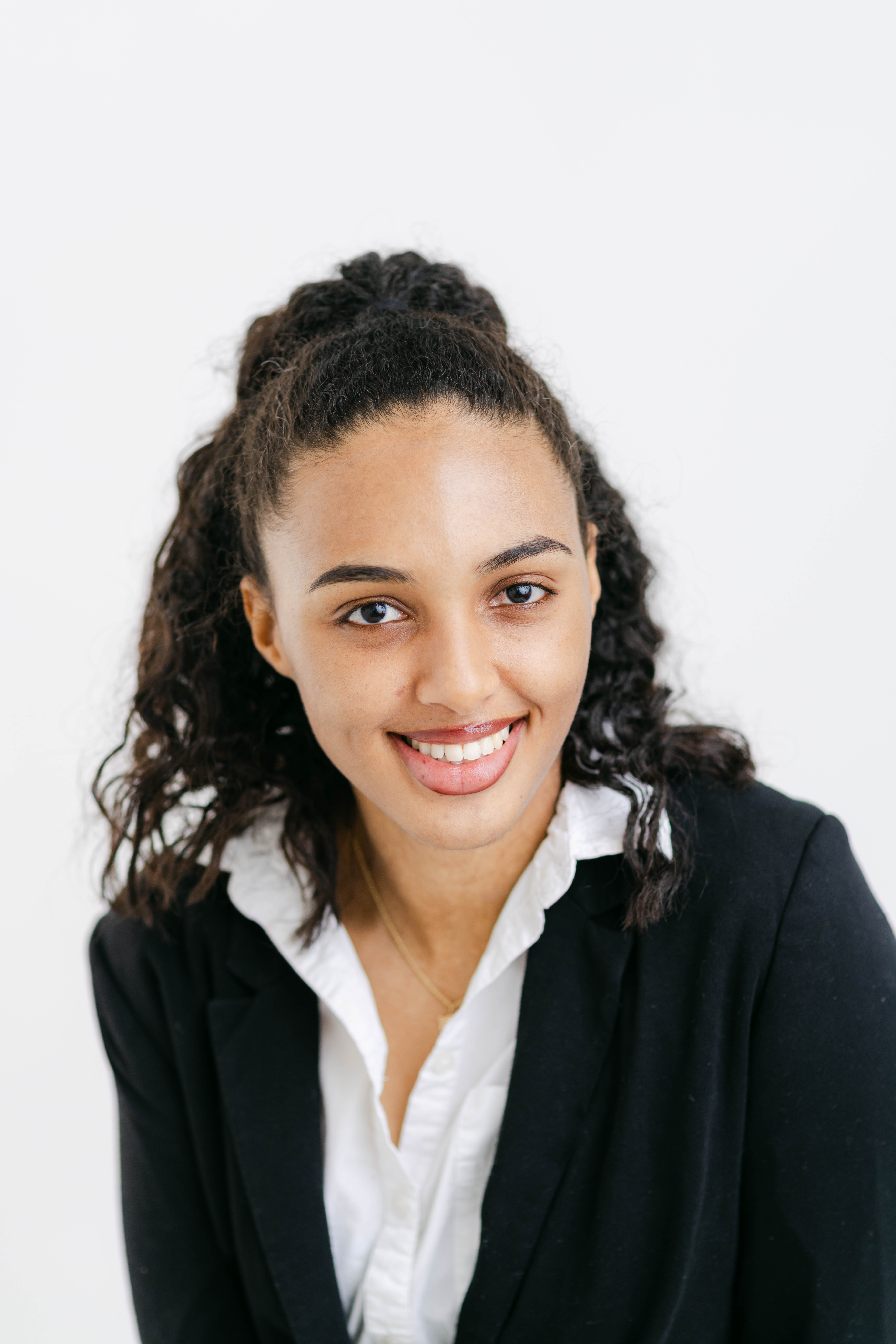}}]{Savana J. Ammons}
received a B.S. in mathematics from Harvey Mudd College in Claremont, CA, USA in 2020, and a Ph.D. in mathematics from the University of Illinois Urbana-Champaign in Champaign, IL, USA in 2025, where she studied partial differential equations involving water waves. She is currently a computational mathematician at Mobius Logic Inc., where she specializes in multi-agent reinforcement learning (MARL) algorithms. She has developed a MARL algorithm that uses automatic curriculum learning to encourage novel agent behavior, created a modularized space simulation engine designed to handle physics interactions and frame translations, and constructed a framework for identifying adversarial agent intent using clustering algorithms and decision trees.
\end{IEEEbiography}

\end{document}